\documentclass[review,12pt]{elsarticle}
\usepackage{geometry}
\usepackage{amsmath,natbib,mathrsfs,stmaryrd,titletoc,subfigure}
\usepackage{amssymb,amsfonts,amsthm,stmaryrd}
\usepackage{hyperref}
\usepackage{lineno}
\modulolinenumbers[1]
\usepackage{caption}
\usepackage{booktabs}
\usepackage{xcolor}
\usepackage{soul}
\usepackage{tabularx}
\journal{arXiv}

\begin{document}

\begin{frontmatter}
\title{Deep learning-enabled large-scale analysis of particle geometry-lithiation correlations in battery cathode materials}
\author[label1]{Binbin Lin\corref{cor1}}
\author[label2,label3]{Luis J. Carrillo\corref{cor1}}
\author[label1]{Xiang-Long Peng\corref{cor2}}
\ead{xianglong.peng@tu-darmstadt.de}
\author[label1]{Wan-Xin Chen}
\author[label2,label3]{David A. Santos}
\author[label2,label3,label4,label5]{Sarbajit Banerjee\corref{cor2}}
\ead{sarbajit.banerjee@psi.ch}
\author[label1]{Bai-Xiang Xu\corref{cor2}}
\ead{xu@mfm.tu-darmstadt.de}
\address[label1]{Division Mechanics of Functional Materials, Institute of Materials Science, Technischen Universität Darmstadt, Darmstadt 64287, Germany}
\address[label2]{Department of Chemistry, Texas A$\&$M University, College Station, Texas 77842-3012, United States}
\address[label3]{Department of Materials Science and Engineering, Texas A$\&$M University, College Station, Texas 77843-3003, United States}
\address[label4]{Laboratory for Inorganic Chemistry, Department of Chemistry and Applied Biosciences, ETH Zurich, Vladimir-Prelog-Weg 2, CH-8093 Zürich, Switzerland}
\address[label5]{Laboratory for Battery Science, PSI Center for Energy and Environmental Sciences, Paul Scherrer Institute, Forschungsstrasse 111, CH-5232 Villigen PSI, Switzerland}        
\cortext[cor1]{These authors contributed equally to this work.}
\cortext[cor2]{Corresponding author(s).}

\begin{abstract}
A deep learning model is employed to address the challenging problem of V$_2$O$_5$ nanoparticle segmentation and the correlation between the chemical composition and the geometrical features of lithiated V$_2$O$_5$ nanoparticles as an exemplar of a phase-transforming battery cathode material. First, the deep learning-enabled segmentation model is integrated with the singular value decomposition technique and a spectral database to generate accurate composition and phase maps capturing lithiation heterogeneities as imaged using scanning transmission X-ray microscopy. These phase maps act as the output properties for correlation analysis. Subsequently, the quantitative influences of the geometrical features of nanoparticles such as the particle size (i.e., projected perimeter and area), the aspect ratio, circularity, convexity, and orientation on the lithiation phase maps are revealed. These findings inform strategies to improve lithiation uniformity and reduce stress in phase-transforming lithium battery materials via optimized particle geometry.
\end{abstract}


\end{frontmatter}

\section{Introduction}
Lithium-ion (Li-ion) batteries have achieved ubiquity in consumer electronics and are rapidly expanding into  electric vehicles and large-scale grid storage systems. Consequently, there is intense interest in increased performance, resilience, and longevity of these systems even under demanding operational conditions \cite{Whittingham2020,cano2018batteries}. 
However, performance degradations arising from various mechanisms such as dendrite formation, active material loss, and interphase formation resulting in increasing internal resistance continue to be vexatious obstacles \cite{Waldmann2016,kim2019lithium}. These degradation mechanisms often stem from complex electrochemistry-mechanics interactions that span from the atomistic scale to the scale of entire porous electrode architectures \cite{kim2019lithium,li2021peering,bai2021chemo,xing2022unraveling,chen2025fracture}. 

For instance, solid active materials in electrodes are subjected to large volumetric changes during cycling, resulting in stress accumulation, and ultimately particle fracture or delamination \cite{cogswell2012coherency,xu2016phase,zhao2019review,Santos2022}, which is strongly  related to lithiation processes and their heterogeneities therein. In electrodes, active materials are typically embedded in the form of particles with varying degrees of agglomeration spanning the range from "single crystals" to "meatball" configurations. Across this entire range, the particle geometries profoundly modify lithiation heterogeneity, specifically patterns of lithiation/delithiation, such as the directionality of intercalation fronts or surface nucleation patterns \cite{fraggedakis2020scaling}. However, localized lithiation processes and their coupling with particle geometry remain largely unexplored \cite{yang2020three,Luo2022}.

Probing and understanding lithiation processes requires adequate resolution and chemical composition sensitivity across length scales. State-of-the-art X-ray spectromicroscopy and electron microscopy methods show promise for probing the aforementioned processes using various contrast methods \cite{wolfman2017visualization,Yuan2017}.  X-ray spectromicroscopy utilizes highly tunable X-rays from a synchrotron source to map pixel/voxel-wise X-ray absorption features across a sample \cite{wolfman2017visualization}, which enables visualization of different compositions and critical atomistic/geometrical correlations underpinning dynamical intercalation phenomena \cite{cogswell2018size,santos2020bending,lim2016origin}. When compared to electron microscopy, X-ray methods offer a large amount of tunability and high penetration power while simultaneous minimizing sample damage during data collection \cite{wolfman2017visualization,du2018relative}. 

Typically, X-ray spectromicroscopy experiments involve transmission intensities that are recorded within a predefined field of view over a single or various X-ray absorption edges. As such, this technique provides spatially resolved chemical composition with element and orbital specificity in addition to probing modification of particle topologies \cite{hitchcock2015soft}. A representative dataset from X-ray spectromicroscopy can contain up to millions of individual spectra and hundreds of images, capturing complex chemical information about a material while simultaneously providing geometrical images with nanometer-scale spatial resolution. Thus, a challenge that must be addressed is the extraction of chemically intuitive comprehension of heterogeneous particles and the link between the dimension of those particles and the respective geometries \cite{Santos2022,Luo2022}.  Machine learning and statistical regression techniques are promising for more effective utilization of such datasets in order to correlate observed dynamical evolution of lithiation inhomogeneities to geometrical features. On the other hand, it is critical that such approaches derive from a physically meaningful interpretation of compositional data from datasets such as X-ray absorption spectra \cite{roychoudhury2023efficient}.

In this work, a deep learning model based on the Mask R-CNN algorithm \cite{lin2022deep} is employed to overcome the challenges associated with instance segmentation and subsequent geometry-chemistry coupling in the material system V$_2$O$_5$. Orthorhombic layered $\alpha$-V$_2$O$_5$ is a thermodynamically stable polymorph of vanadium pentoxide with multiple accessible redox states (V$^5+$/V$^4+$ and V$^4+$/V$^3+$) and an atomistic framework that can incorporate up to 3 Li ions per formula unit (i.e., up to a final composition of Li$_3$V$_2$O$_5$). This material has drawn much attention in the battery community due to its high theoretical capacity and its ability to readily change redox states \cite{santos2021assessing}. From an instance segmentation perspective, the V$_2$O$_5$ nanoparticle dispersions featured in this work, combined with their non-spherical geometry, lead to considerable irregularity in thickness, length-to-width aspect ratios, and edge profiles, which collectively result in significant variability in particle shape despite the all-encompassing nanowire classification. This shape variability combined with particle agglomeration makes for an ambitious instance segmentation task that closely resembles the needs of many applications today characterized by highly polydisperse particle geometries \cite{monchot2021deep,mansfeld2019towards}. Through the lens of chemistry-mechanics coupling, V$_2$O$_5$ appears as a fascinating case study for the geometry-lithiation pattern correlation analysis as a binary transition metal oxide cathode manifesting several possible lithiation-induced phase transitions that exacerbate heterogeneities arising from diffusion limitations \cite{santos2020bending,andrews2020curvature,horrocks2013finite}. 

 After developing a robust instance segmentation model, the single value decomposition (SVD) technique in conjunction with a curated spectral database \cite{Santos2022} is applied to obtain highly accurate quantitative compositional lithiation phase maps of particle ensembles. These compositional phase maps are then used as the output property for correlation analysis of the particle descriptors segmented from the same dataset. The correlation analysis reveals the relationships between the geometrical features and the actual phase maps, and identifies how each geometrical feature affects the property of interest. The lithiation patterns of the nanoparticles are found to strongly depend on the particle geometries. This study contributes insightful guidelines on how to enhance the chemical performance of V$_2$O$_5$-based lithium battery electrode materials via engineering the particle geometries.
\section{Material and methods}
\subsection{Synthesis, lithiation and delithiation of V$_2$O$_5$ nanoparticles}
Two sets of V$_2$O$_5$ nanoparticle samples are considered. The first set of samples with broad variations in geometry features is derived from the same synthetic methodology investigated in our previous work \cite{lin2022deep}. In that work, only the nanoparticle segmentation is considered. Here, the lithiation phase maps are extracted and analyzed. Details on the synthesis of these samples are found in \cite{lin2022deep}. This set of samples is employed for the geometry-lithiation correlation analysis. 

The second set of nanoparticle samples newly introduced in this work exhibits more rounded particle shapes in comparison with the first set of samples. These samples are taken to demonstrate the generality of the proposed method, i.e., it is applicable to different types of nanoparticles. The details on the synthesis of these samples and corresponding results are found in the Supplementary Material. Since the geometry variations of these nanoparticles are narrow, the geometry-lithiation correlation analysis on them is not feasible and hence is not investigated.  
\subsection{Mask R-CNN model}
\label{ML}
The Mask R-CNN ML model used for particle segmentation in this work is originally proposed and trained in our previous work \cite{lin2022deep}. The model includes three main parts: feature extraction, region of interest proposal, and mask and bounding box prediction and classification. In the feature extraction stage, the feature maps of input images are extracted by utilizing multiple CNN layers. In the region of interest proposal stage, based on the feature maps from the previous stage, the region proposal network is exploited to propose a region of interest with a known confidence score for each particle, which is then modified and aligned to have a uniform size. Finally, in the mask and bounding box prediction and classification stage, the prediction heads classify the object, predict bounding boxes, and generate binary instance masks for particle segmentation. The model is trained by synthetic data, and its performance has been verified by experimental data. More details on the model structure, hyperparameters, and training and evaluation are found in \cite{lin2022deep}.
\section{Results and Discussion}
\begin{figure}[htbp]
	\centering
	\includegraphics[width=0.95\textwidth]{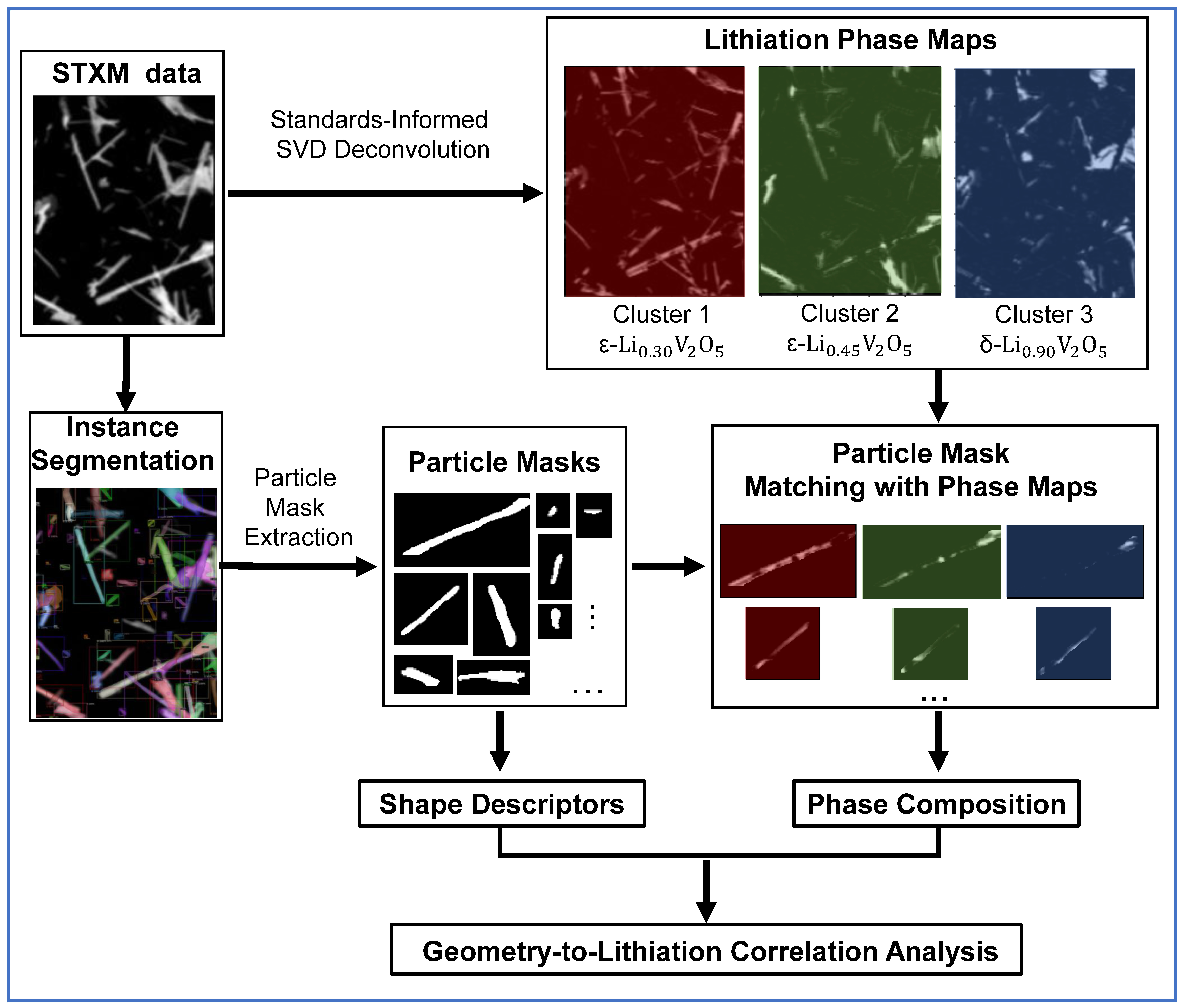}
	\caption{Overview of the workflow for geometry-lithiation correlation analysis based on STXM data. }
	\label{fig:workflow}
\end{figure}

Based on the developed deep learning model for particle segmentation in Section \ref{ML}, the correlations of the geometrical descriptors obtained from the segmentation results with the lithiation patterns presented for the STXM data are investigated here. The general workflow is shown in Fig.\,\ref{fig:workflow}. 
\subsection{Deep learning-enabled particle segmentation}
 The representative dataset is a network ensemble of more than 80 polydisperse lithiated V$_2$O$_5$ particles, shown in the STXM figure in Fig.\,\ref{fig:workflow}. This dataset demonstrates the need to use advanced image analysis tools due to the significant number of particles and the overlap among them. The dataset is first processed by SVD to obtain distinctive lithiation phase maps. This results in three distinct patterns that accurately capture visualizations of the spatial distribution of lithiation pattern clusters 1-3, corresponding to $\varepsilon$-Li$_{0.3}$V$_2$O$_5$, $\varepsilon$-Li$_{0.45}$V$_2$O$_5$, and $\delta$-Li$_{0.9}$V$_2$O$_5$ phases (Fig.\,\ref{fig:workflow}), respectively. Meanwhile, the previously trained instance segmentation ML model (Section \ref{ML}) is applied to obtain the masks and bounding boxes of individual nanowire particles in the STXM image. The individual masks reflect the projected contour and shape of individual particles and form the working basis for further morphological feature extraction. They are stored in a multidimensional array of size $H \times W \times N$, where $H$  and $W$ are the height and width of the STXM image, and $N$ is the number of segmented particles. In Fig.\,\ref{fig:workflow}, the binary masks are cropped to the size of the particle for better visualization. 
\subsection{Geometrical descriptors and their statistics across nanoparticle ensembles}
Using the \textit{findContours} function from the \textit{OpenCV} library, the contour of the binary particle mask is extracted and used to compute the geometrical descriptors for each particle. 
The following 8 descriptors that quantify the morphological characteristics of the particle shapes are defined as follows:
\begin{itemize}	
	\item[$\bullet$] Projected particle perimeter $P$ is defined as the length of the contour boundary.
	\item[$\bullet$] Projected particle area $A$ refers to the number of pixels enclosed within the contour boundary.
	\item[$\bullet$] Aspect ratio is defined as the ratio of the long edge $L_\text{l}$ to the short edge $L_\text{s}$ of the detected bounding box, i.e., $L_\text{l}/L_\text{s}$ .
	\item[$\bullet$] Circularity $4 \pi A / P^2$ measures the degree to which a shape approximates a perfect circle. Its value is 1 for a perfect circle and decreases as the shape becomes less circular.
	\item[$\bullet$] Eccentricity is defined as the square root of the squared difference of the major axis $L_\text{M}$ and the minor axis $L_\text{m}$ divided by the major axis when an ellipse is fitted to the particle shape, i.e., $\sqrt{L_{\text{M}}^2 - L_{\text{m}}^2}/L_\text{M}$. The key difference between eccentricity and aspect ratio lies in the reference shape used for fitting: aspect ratio is based on the lengths of the long and short edges of a fitted bounding rectangle, whereas eccentricity is based on the major and minor axes of a fitted ellipse.  Eccentricity provides a more direct measure of elongation when the particle shape more closely resembles an ellipse than a rectangle.
	\item[$\bullet$] Convexity is defined as the ratio of the convex perimeter $P_\text{covex}/P$ to the actual perimeter $P$ of a shape, i.e., $P_\text{covex}/P$. The convex perimeter is obtained by fitting a convex hull around the shape, i.e., the smallest convex polygon enclosing all points of the non-convex shape. Convexity serves as an indicator of the shape irregularity (e.g., surface roughness).
	\item[$\bullet$] Solidity is defined as the area $A$ divided by the convex hull area $A_\text{covex}$, i.e., $A/A_\text{covex}$. It describes the extent to which a shape is convex or concave~\cite{zdilla2016circularity}.
	\item[$\bullet$] Orientation is defined as the angle $\theta$ between the major axis of the ellipse fitted to the particle and the horizontal axis.
\end{itemize}
Note that the projected particle perimeter and area are particle size-dependent and thus represent absolute measures, whereas the remaining six descriptors are relative, as they are particle size-independent. The considered descriptors and their statistical values are summarized in Table\,\ref{tab:Descriptor_battery}. The mean and variance numbers provide initial insight into the variety in the morphology of particles in the extracted dataset. As can be seen in Table \ref{tab:Descriptor_battery}, larger numbers in perimeter, area, and aspect ratio deviation indicate a stronger variation in particle size distribution and morphology in the particle ensembles. The smaller number in circularity indicates that the particles have a significant deviation from circular shapes, but with relatively smooth boundaries indicated by the high convexity, which can be visually confirmed by the generally nanowire-shaped particles. The low circularity value indicates that the particles deviate significantly from a circular shape. While the high convexity value suggests that their boundaries are relatively smooth. These are consistent with the visual observation of their generally nanowire-like morphology.
\begin{table}[] 
	\caption{Geometrical descriptors, their definition and statistical numbers for the presented dataset.}
	\centering
	\begin{tabular}{c c c c} 
		\hline
		Descriptor                 & Definition                       & Mean  &       Std.    \\
		\hline
		Projected particle perimeter & $P$ [Pixels]                            & 84.5   &     74.6     \\
		Projected particle area      & $A$ [Pixels]                            & 222.4  & 275.9        \\
		Aspect ratio                 & $L_\text{l}/L_\text{s}$ [-]                    &  3.4 &  2.2             \\
		Circularity                  & $4 \pi A / P^2$ [-] & 0.42 & 0.21                           \\
        Eccentricity                 & $\sqrt{L_{\text{M}}^2 - L_{\text{m}}^2}/L_\text{M}$ [-]   & 0.89        &    0.16  \\		
		Convexity                    & $P_\text{covex}/P$  [-]                   & 0.95 & 0.03          \\
		Solidity                     &  $A/A_\text{covex}$ [-]          &        0.83 & 0.11            \\
		Orientation                  & $\theta$       &  89  & 56 \\
		\hline                                            
	\end{tabular}
	\label{tab:Descriptor_battery}
\end{table}
\subsection{Phase composition of extracted nanoparticles}
\label{phase_composition}
Denote the binary masks as $\mathcal{M}^N_{hw}$ for each $N$-th mask, and the deconvoluted phase maps as $\phi^i_{hw}$ ($i = 1,\, 2,\, 3$) for each $i$-th cluster. The composition map for each particle is the superposition of each phase map weighted by its corresponding stoichiometry coefficient \cite{andrews2020curvature}, i.e., 
\begin{equation}\label{eq:phase_composition}
	C^N_{hw} = \sum_i x^i\mathcal{M}^N_{hw}\phi^i_{hw}
\end{equation}
with $x^1 = 0.3$, $x^2 = 0.45$, and $x^3 = 0.9$ being the stoichiometry coefficients of the phase clusters and the lower subscript $hw$ denoting the pixel position in the image. Using Eq.\,\eqref{eq:phase_composition}, the compositional percentage of each contributing phase to the total composition can be determined as
\begin{equation}\label{Eq:Composition}
	CP^{N,\,i} = \sum_h\sum_w \frac{x^i\mathcal{M}^N_{hw}\phi^i_{hw}}{C^N_{hw}}.
\end{equation}

\subsection{Particle geometry-lithiation correlations}
\begin{figure}[htbp]
	\centering         \includegraphics[width=0.95\columnwidth]{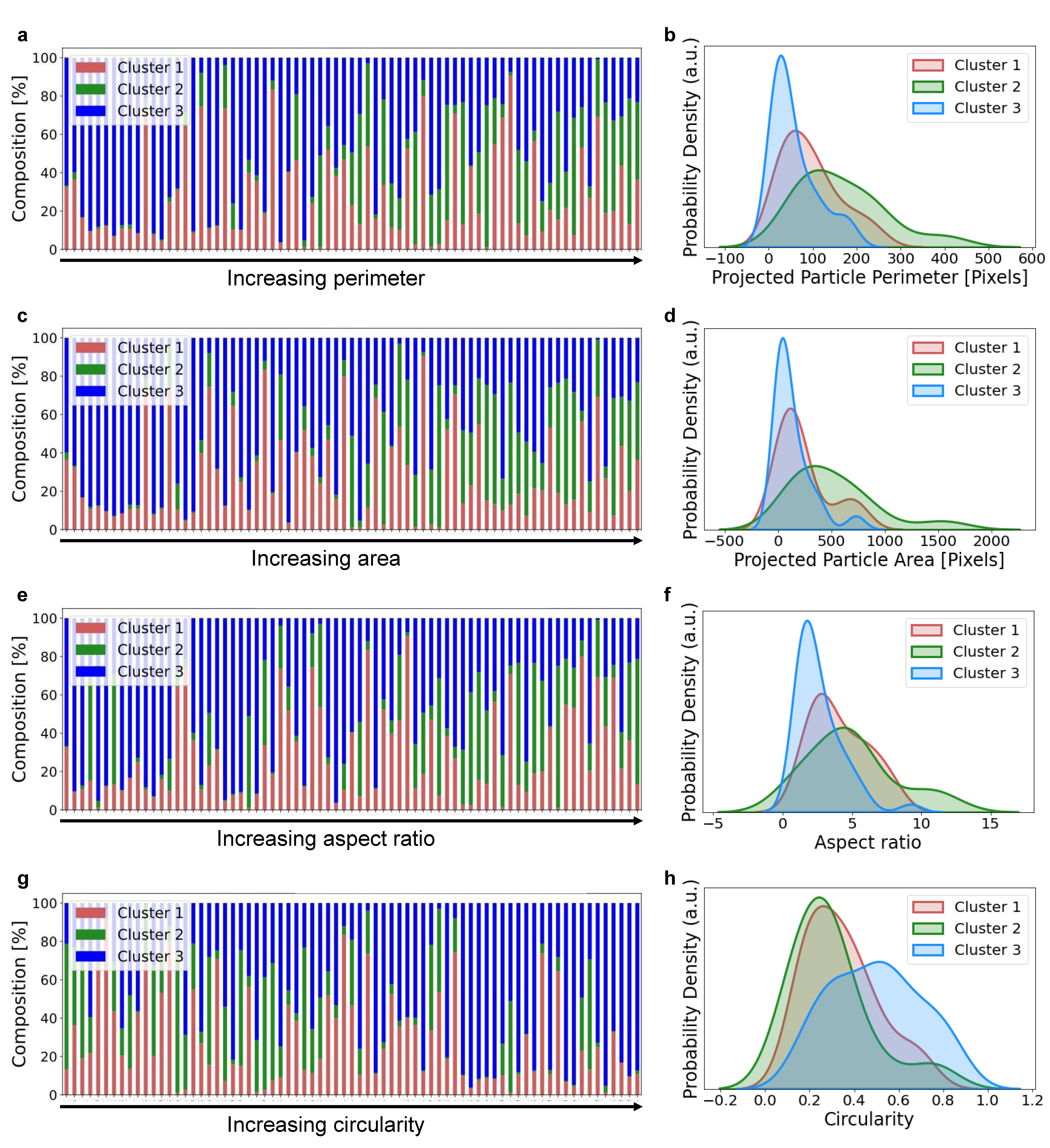}
	\caption{Stacked bar plot of particle composition ($CP^{N,\,i}$) and the KDE curves of the three clusters for different descriptors: (a-b) Projected particle perimeter, (c-d) Projected particle area, (e-f) Aspect ratio, and (h-g) Circularity. The four descriptors show clear correlations with lithiation state. Higher perimeter, area, and aspect ratio yield lower lithiation levels and increased phase segregation. In contrast, higher circularity corresponds to higher lithiation levels and less phase segregation. This trend may be affected by particle size variation.} 
	\label{fig:perimeter_cluster}
\end{figure}

\begin{figure}[htbp]
	\centering
	\includegraphics[width=0.95\columnwidth]{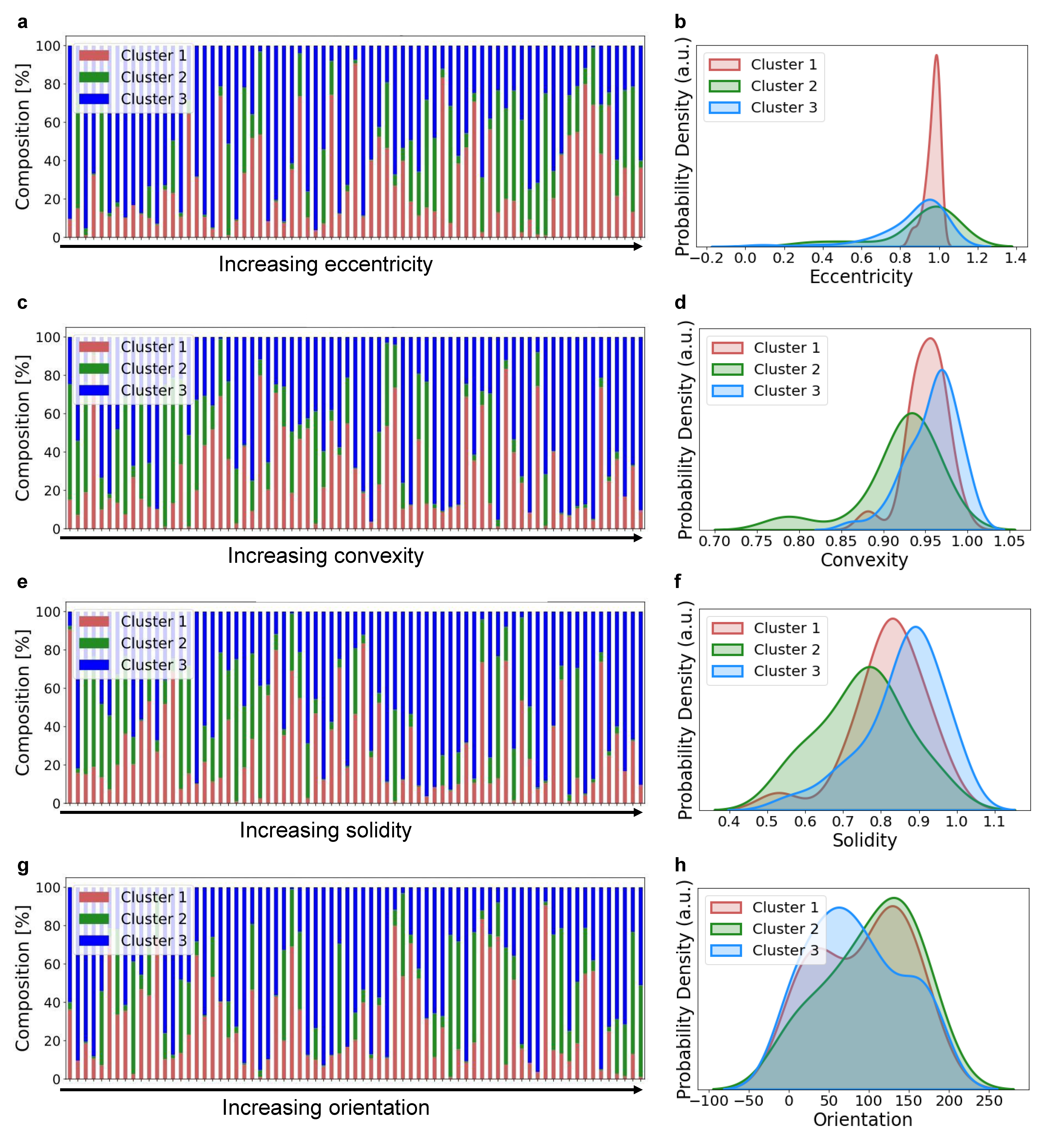}
		\caption{Stacked bar plot of particle composition ($CP^{N,\,i}$) and the KDE curves of the three clusters for different descriptors: (a-b) Eccentricity, 
			(c-d) Convexity, (e-f) Solidity, and (h-g) Orientation. Eccentricity shows an inverse correlation with lithiation state. Particles of higher eccentricity tend to exhibit lower lithiation level and increased phase segregation. In contrast, convexity and solidity exhibit a direct correlation. No clear trend is observed for orientation, implying that its influence on lithiation behavior can be neglected.} 
	\label{fig:convexity_cluster}
\end{figure}

The method introduced in Section\,\ref{phase_composition} is exploited to evaluate the phase compositions of all particles in the STXM dataset. The corresponding results are found in Figs.~\ref{fig:perimeter_cluster} and \ref{fig:convexity_cluster}. The sub-figures in the left column show the plots of particle composition versus segmented particle IDs arranged by ascending descriptor values. This visualization serves two purposes. First, it allows for assessing the correlation between descriptors and the state of lithiation, i.e., whether a particle is highly or relatively less lithiated indicated by the cluster percentages (e.g., a high percentage of cluster 1 and that of cluster 3 indicate a low lithiation state and a high lithiation state, respectively). Second, it enables visualizing phase segregation, i.e., the extent to which different phases coexist in a single particle. To quantitatively evaluate the relationships between lithiation states and descriptors, the right column presents kernel density estimation (KDE) curves of statistical distribution of the descriptors for each cluster. From Figs.\,\ref{fig:perimeter_cluster} and \ref{fig:convexity_cluster}, the lithiation state is strongly correlated with the descriptors.  

First, a clear inverse correlation is observed between the projected 2D particle perimeter and area and the lithiation level in Figs.\,\ref{fig:perimeter_cluster}a and \ref{fig:perimeter_cluster}c. As the particle perimeter and area increase, the blue bars representing Cluster 3 (high lithiation state) shrink and the lithiation pattern shifts towards less lithiated Clusters 1 (red) and 2 (green), indicating the decrease of lithiation level. The KDE plots in Figs.\,\ref{fig:perimeter_cluster}b and \ref{fig:perimeter_cluster}d also support this observed trend.  Cluster 3 exhibits a peak density at a perimeter of approximately 20-30 pixels and an area of around 80-100 pixels. In contrast, the peak densities of Clusters 1 and 2 shift more noticeably toward larger perimeter and area values.
Furthermore, as illustrated by the mixed composition of stacked bars in Figs.\,\ref{fig:perimeter_cluster}a and \ref{fig:perimeter_cluster}c, phase segregation becomes more pronounced with increasing particle area and perimeter. A similar inverse correlation is observed for the aspect ratio, as shown in Figs.\,\ref{fig:perimeter_cluster}e and \ref{fig:perimeter_cluster}f. Higher aspect ratios are associated with reduced lithiation levels and increased phase segregation.

Circularity quantifies how much a particle shape deviates from a perfect circle. For the nanowire-like particles considered here, circularity values are relatively low, ranging from 0.2 to 0.3 for Clusters 1 and 2, and between 0.4 and 0.6 for the more broadly distributed Cluster 3, as shown in the distribution plot in Fig.\,\ref{fig:perimeter_cluster}h. This suggests that particles with more circular shapes tend to exhibit higher lithiation levels and weaker phase segregation according to Fig.\,\ref{fig:perimeter_cluster}g. However, this interpretation should be treated with caution, as the broad size variation among the particles may influence circularity since this shape descriptor is not evaluated on the same size basis.

Eccentricity characterizes how closely a particle shape resembles an ellipse. As shown in Fig.\ref{fig:convexity_cluster}a, an inverse correlation is observed between eccentricity and lithiation level, with a higher eccentricity corresponding to a lower lithiation level. Notably, while the eccentricity value corresponding to the peak density is high for all clusters, the peak density is particularly pronounced for cluster 1 as shown in Fig.\,\ref{fig:convexity_cluster}b. These results suggest that highly eccentric particles tend to be less lithiated. This finding is consistent with the observations of Mistry et al.\,\cite{mistry2022asphericity}, who reported that phase segregation or heterogeneity in lithium distribution within particles is at least partially driven by geometrical asphericity (i.e., high eccentricity, high aspect ratio) and the presence of extended defects that disrupt Li-ion diffusion pathways.

The density distributions of convexity and solidity show relatively small variation across the three clusters, reflecting the overall similarity in particle morphology despite significant differences in particle size. Nonetheless, both descriptors exhibit a direct correlation with the lithiation state, as indicated by the increasing proportion of blue bars with higher convexity and solidity values Figs.\,\ref{fig:convexity_cluster}c and \ref{fig:convexity_cluster}e. This trend aligns with that reflected by the density distribution plots in Figs.\,\ref{fig:convexity_cluster}d and \ref{fig:convexity_cluster}f. From Figs.\,\ref{fig:convexity_cluster}g and \ref{fig:convexity_cluster}h, there is no clear correlation observed for the orientation descriptor. It can be inferred that its influence on the property of interest is insignificant.

Based on the above discussion, reducing particle size and perimeter appears to increase the lithiation level. While, increasing shape descriptors such as circularity, convexity, and solidity tends to enhance lithiation and reduce phase segregation. However, the observed trend of decreasing perimeter with increasing lithiation, as shown in Fig.\,\ref{fig:perimeter_cluster}a, should be interpreted with caution. This trend is largely influenced by the considerable variation in particle size within the ensemble. In particular, smaller particles are generally more highly lithiated, and naturally, smaller particle areas are associated with smaller perimeters. 
\begin{figure}[htbp]
	\centering
	\includegraphics[width=0.75\columnwidth]{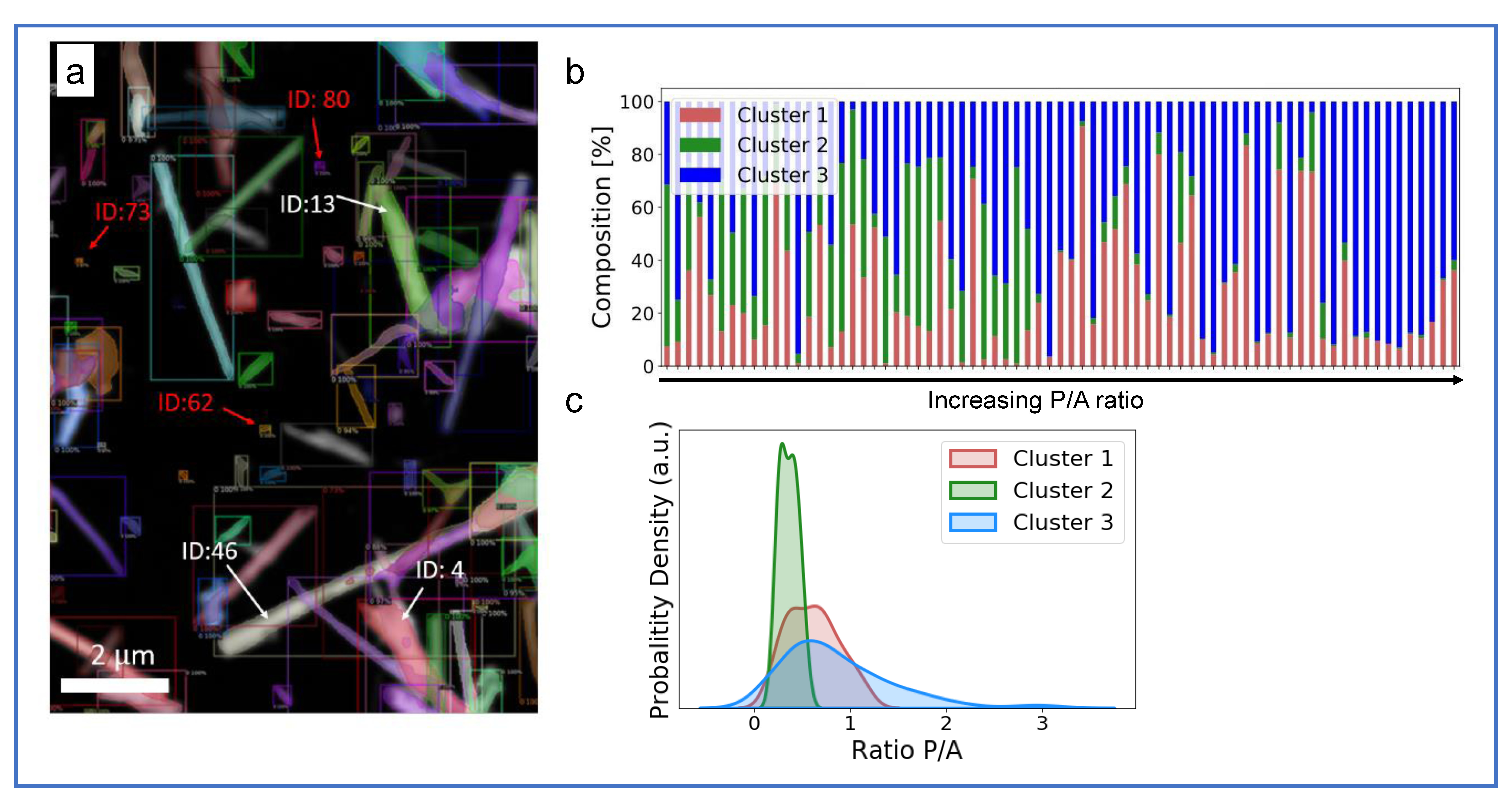}
	\caption{ Influence of the additional descriptor, i.e., P/A ratio: (a) STXM image of instance segmented particles with particles of interest and their IDs highlighted, (b) Stacked bar plot of particle composition ($CP^{N,\,i}$), and (c) KDE curves of the three clusters. In (a), red IDs 62, 73, and 80 point to the particles with the largest P/A ratio. White IDs 4, 13, and 46 point to the particles with the smallest P/A ratio. From (b), particles of low P/A ratio show a high degree of phase coexistence, whereas particles of high P/A ratio have a more homogeneous concentration and a higher lithiation level. This trend is supported by the KDE curves in (c). }
	\label{fig:P-A_ratio}
\end{figure}

To address the limitations associated with particle size variation, an additional descriptor is introduced, i.e., the perimeter-to-area (P/A) ratio, which can be viewed as an analog to the surface-area-to-volume ratio (or specific surface area) in a 3D context. This descriptor is widely used in scientific and engineering applications involving diffusion or adsorption, where transport occurs through the surface \cite{planinvsivc2008surface}. The representative particles with low and high P/A ratios are illustrated in Figs.\,\ref{fig:P-A_ratio}a. The correlation results are shown in Figs.\,\ref{fig:P-A_ratio}b and \ref{fig:P-A_ratio}c.  In the present context, a higher P/A ratio implies more surface area per unit volume for lithium-ion diffusion, potentially enabling faster uptake and leading to a higher lithiation state, assuming a constant diffusion rate within the particle. This trend is evident in Figs.\,\ref{fig:P-A_ratio}b and \ref{fig:P-A_ratio}c, where increasing the P/A ratio corresponds to a greater proportion of Cluster 3 (blue), and hence a higher lithiation level. Therefore, the P/A ratio emerges as the key geometrical descriptor, whereas particle size or perimeter alone may be insufficient to explain the lithiation behavior. 

The above analysis of phase coexistence and its relationship with particle geometries offers valuable insights into the optimal design of the constituent particles of battery electrode architectures. Reducing particle dimensions can weaken phase segregation \cite{Santos2022} and hence mitigate interfacial mismatch stresses induced by phase boundaries. Indeed, Luo et al. \cite{Luo2022} demonstrated that micrometer-sized V$_2$O$_5$ platelets exhibit substantial variation in phase distribution across electrode thickness, whereas electrodes composed of homogeneous, nanometer-sized spherical V$_2$O$_5$ particles show uniform lithiation behavior. Furthermore, based on the isoperimetric inequality, which states that a sphere has the minimal surface-to-volume ratio for a given volume, a morphological design strategy can be proposed: introducing sharp features or spike-like structures to increase the specific surface area, thereby enhancing lithium uptake and improving lithiation levels.

\section{Conclusion}
In this work, we presented a deep learning-based framework for particle segmentation in STXM datasets and established quantitative correlations between nanoparticle morphology and lithiation behavior of V$_2$O$_5$ nanoparticles. The developed instance segmentation model enabled the extraction of individual particle masks from complex, polydisperse structures, forming the basis for detailed morphological analysis using a set of geometrical descriptors. By coupling these descriptors with deconvoluted lithiation phase maps via pixel-wise operations, we quantified the phase composition of each particle and analyzed its dependence on the geometrical descriptors.

Our results demonstrate that particle geometry significantly influences the lithiation pattern. In particular, smaller particles with lower projected area and perimeter, as well as those with lower aspect ratio and eccentricity, tend to exhibit higher lithiation states and reduced phase segregation. Similarly, increased circularity, convexity, and solidity lead to more homogeneous lithiation. Particularly, the P/A ratio is an important metric capturing the surface-to-volume effect relevant for lithium diffusion in particles. These findings are consistent with some existing experimental studies (e.g., \cite{Santos2022,Luo2022}). 

The proposed methodology provides a powerful tool for linking particle-scale morphology to electrochemical behavior in battery materials and offers new insights for the microstructural design of high-performance electrodes.

\section*{Acknowledgments}
This work is supported by German Research Foundation (DFG). B. Lin, B-X. Xu acknowledge the financial support under the grant agreement No. 405422877 of the Paper Research project (FiPRe) and the Federal Ministry of Education and Research (BMBF) and the state of Hesse as part of the NHR Program. The authors gratefully acknowledge the computing time granted by the NHR4CES Resource Allocation Board and provided on the supercomputer Lichtenberg II at TU Darmstadt as part of the NHR4CES infrastructure. The calculations for this research were conducted with computing resources under the project project1020, special0007.  A portion of the STXM measurements utilized in this work was collected at the Canadian Light Source, which is supported by the Natural Sciences and Engineering Research Council of Canada, the National Research Council Canada, the Canadian Institutes of Health Research, the Province of Saskatchewan, Western Economic Diversification Canada, and the University of Saskatchewan. The research at Texas A\&M University was primarily supported by the National Science Foundation under
DMR 1809866. Use of the TAMU Materials Characterization Facility (RRID: SCR\_022202) is acknowledged. D.A.S. acknowledges support under a NSF Graduate Research Fellowship under grant No. 1746932.

\section*{Competing Interests}
The authors declare no competing financial or non-financial interests.

\section*{Author contribution}
\textbf{Binbin Lin}: Writing – original draft, Visualization, Methodology, Software, Formal analysis, Data curation, Conceptualization. \textbf{Luis J. Carrillo} : Writing – original draft, Visualization, Methodology, Formal analysis, Data curation, Conceptualization. \textbf{Xiang-Long Peng}: Writing-original draft, Visualization, Formal analysis, Conceptualization. \textbf{Wan-Xin Chen}: Writing – review \& editing, Formal analysis. \textbf{David A. Santos}: Methodology, Data curation. \textbf{Sarbajit Banerjee}: Writing – review \& editing, Supervision, Resources, Project administration, Funding acquisition, Conceptualization. \textbf{Bai-Xiang Xu}: Writing – review \& editing, Supervision, Resources, Project administration, Funding acquisition, Conceptualization.

\bibliographystyle{model3-num-names}

\end{document}


\begin{frontmatter}

\title{Supplementary Material: \\Deep learning-enabled large-scale analysis of particle geometry-lithiation correlations in battery cathode materials}
\author[label1]{Binbin Lin\corref{cor1}}
\author[label2,label3]{Luis J. Carrillo\corref{cor1}}
\author[label1]{Xiang-Long Peng\corref{cor2}}
\ead{xianglong.peng@tu-darmstadt.de}
\author[label1]{Wan-Xin Chen}
\author[label2,label3]{David A. Santos}
\author[label2,label3,label4,label5]{Sarbajit Banerjee\corref{cor2}}
\ead{sarbajit.banerjee@psi.ch}
\author[label1]{Bai-Xiang Xu\corref{cor2}}
\ead{xu@mfm.tu-darmstadt.de}
\address[label1]{Division Mechanics of Functional Materials, Institute of Materials Science, Technischen Universität Darmstadt, Darmstadt 64287, Germany}
\address[label2]{Department of Chemistry, Texas A$\&$M University, College Station, Texas 77842-3012, United States}
\address[label3]{Department of Materials Science and Engineering, Texas A$\&$M University, College Station, Texas 77843-3003, United States}
\address[label4]{Laboratory for Inorganic Chemistry, Department of Chemistry and Applied Biosciences, ETH Zurich, Vladimir-Prelog-Weg 2, CH-8093 Zürich, Switzerland}
\address[label5]{Laboratory for Battery Science, PSI Center for Energy and Environmental Sciences, Paul Scherrer Institute, Forschungsstrasse 111, CH-5232 Villigen PSI, Switzerland}        
\cortext[cor1]{These authors contributed equally to this work.}
\cortext[cor2]{Corresponding author(s).}
\end{frontmatter}

\renewcommand{\thefigure}{S.\arabic{figure}}
\renewcommand{\thetable}{S.\arabic{table}}
\renewcommand{\theequation}{S.\arabic{equation}}
\section{Supplementary Material: second set of  V$_2$O$_5$ nanoparticles}
\subsection{Synthesis, lithiation and delithiation}
Bulk V$_2$O$_5$ (99.6\% trace metals basis) was ball-milled for 10 min at $1725\,\text{RPM}$ to produce fine particles. Nanoparticles were synthesized by mixing $1.634\,\text{g}$ of this powder with $0.465\,\text{g}$ oxalic acid (98\%, anhydrous, Thermo Fisher) in $75\,\text{mL}$ of deionized water (conductivity of $18.2\,\text{M/cm}$). This mixture was heated in a laboratory
oven (VWR Gravity Convection Oven, 89, 511–412) at $210\,^\circ\text{C}$ for $72\,\text{h}$. After cooling, the resulting dark green V$_3$O$_7$·H$_2$O precipitate was washed with water and 2-propanol (99\% purity, VWR), then dried under ambient conditions. The product was annealed at $450\,^\circ\text{C}$ for $24\,\text{h}$ in air to reoxidize it to V$_2$O$_5$.

Lithiation of V$_2$O$_5$ particles was performed using $n$-butyllithium ($2.5\,\text{M}$ in hexanes, Sigma, Aldrich) diluted to $0.5\,\text{M}$ in $n$-hexane (DriSolv, $\geq$ 95\%, Sigma, Aldrich) for $10\,\text{s}$ to promote phase segregation while preventing Li saturation and irreversible phase transformations. Delithiation was performed using a $0.5\,\text{M}$ solution of nitrosonium tetrafluoroborate (98\%, Thermo Fisher) dissolved in acetonitrile (DrySolv, $\geq$ 95\%, Sigma, Aldrich) for $10\,\text{s}$. Copious washes of $n$-hexane and acetonitrile were done to ensure $n$-butyllithium and NOBF$_4$ do not mix. 

STXM measurements were performed at the 10ID-1 beamline of the Canadian Light Source in Saskatoon, SK, CA and the 5.3.2.2 beamline at Lawrence Berkeley National Laboratory in Berkeley, CA, USA. The 10ID-1 beamline is equipped with an elliptically polarized Apple II type undulator, which provides an intense beam in the $130–2700\,\text{eV}$ energy range whereas the 5.3.2.2 beamline is equipped with 6 elliptically polarizing inductors, providing an intense beam in the $250–780\,\text{eV}$ energy range. State-of-the-art Fresnel zone plate optics are utilized in conjunction with an order-sorting aperture to achieve a focused X-ray beam. Regions of interest were selected based on experimental goals (e.g., particle size and shape), which are typically identified using a $518\,\text{eV}$ energy (V-L$_3$ peak) to achieve high contrast. Before spectromicroscopy stacks are initiated, a line-scan is typically performed to view average composition and ensure accurate V and O peaks. 

Spectromicroscopy stacks were obtained in the $508–560\,\text{eV}$ energy range corresponding to V-L and O-K absorption edges at a step size of $0.2\,\text{eV}$ in regions of spectral interest while other regions of lower priority (post-edge and pre-edge) were collected with step sizes of $1\,\text{eV}$. Typically, a region of interest is raster-scanned while simultaneously recording transmission intensity in a stepwise fashion by a charged-coupled device (CCD) detector. Image registration and processing were achieved utilizing the cross-correlation analysis function (i.e., Jacobsen stack analyze) in the aXis2000 software suite (Version April 22, 2023). An incident spectrum was extracted from the region pertaining to the transmission of silicon nitride substrate to convert the overall transmission measurements to an optical density matrix. Principal component analysis (PCA) and subsequent $k$-means clustering were implemented utilizing the PCA\_GUI routine (version 1.1.1) within the aXis2000 suite. The number of principal components considered for clustering was chosen based on eigenvalues and their corresponding eigenspectrum/eigenimage representation of the data. Standards published previously aid in preliminary interpretation of eigenspectra, which improves the selection of significant components considerably. The averaged spectrum from each cluster was compared to standards published previously to discern Li stoichiometries for each cluster and subsequently used to obtain phase-specific signatures from which composition maps could be generated by SVD \cite{Santos2022}.
\subsection{Particle segmentation and composition maps}
\begin{figure}[htbp]
	\centering
	\includegraphics[width=0.9\columnwidth]{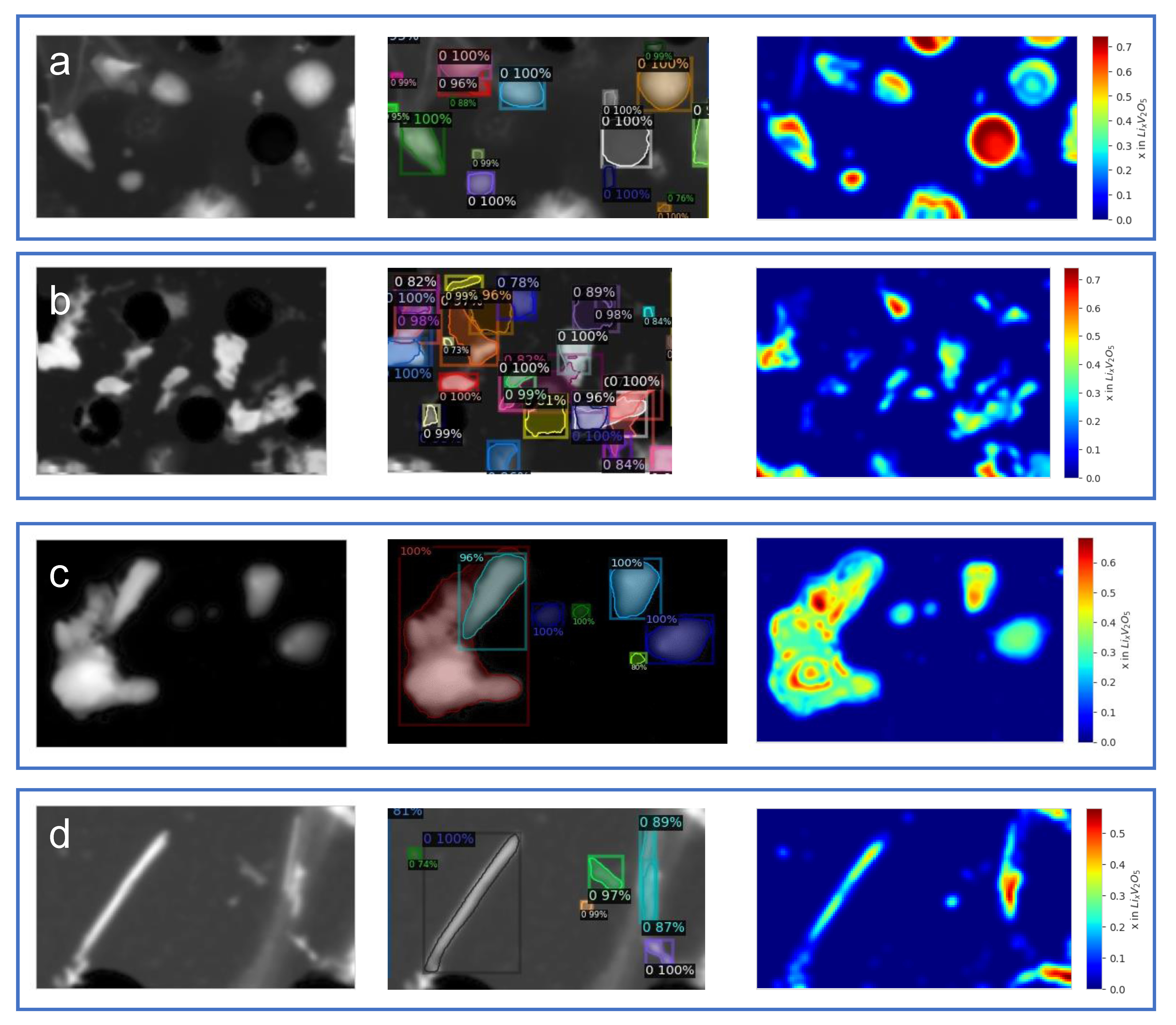}
	\caption{Additional nanoparticle set 1: STXM images (left column), particles after segmentation (middle column), and the lithiation composition maps (right column).}
	\label{fig:set1}
\end{figure}

\begin{figure}[htbp]
	\centering
	\includegraphics[width=0.9\columnwidth]{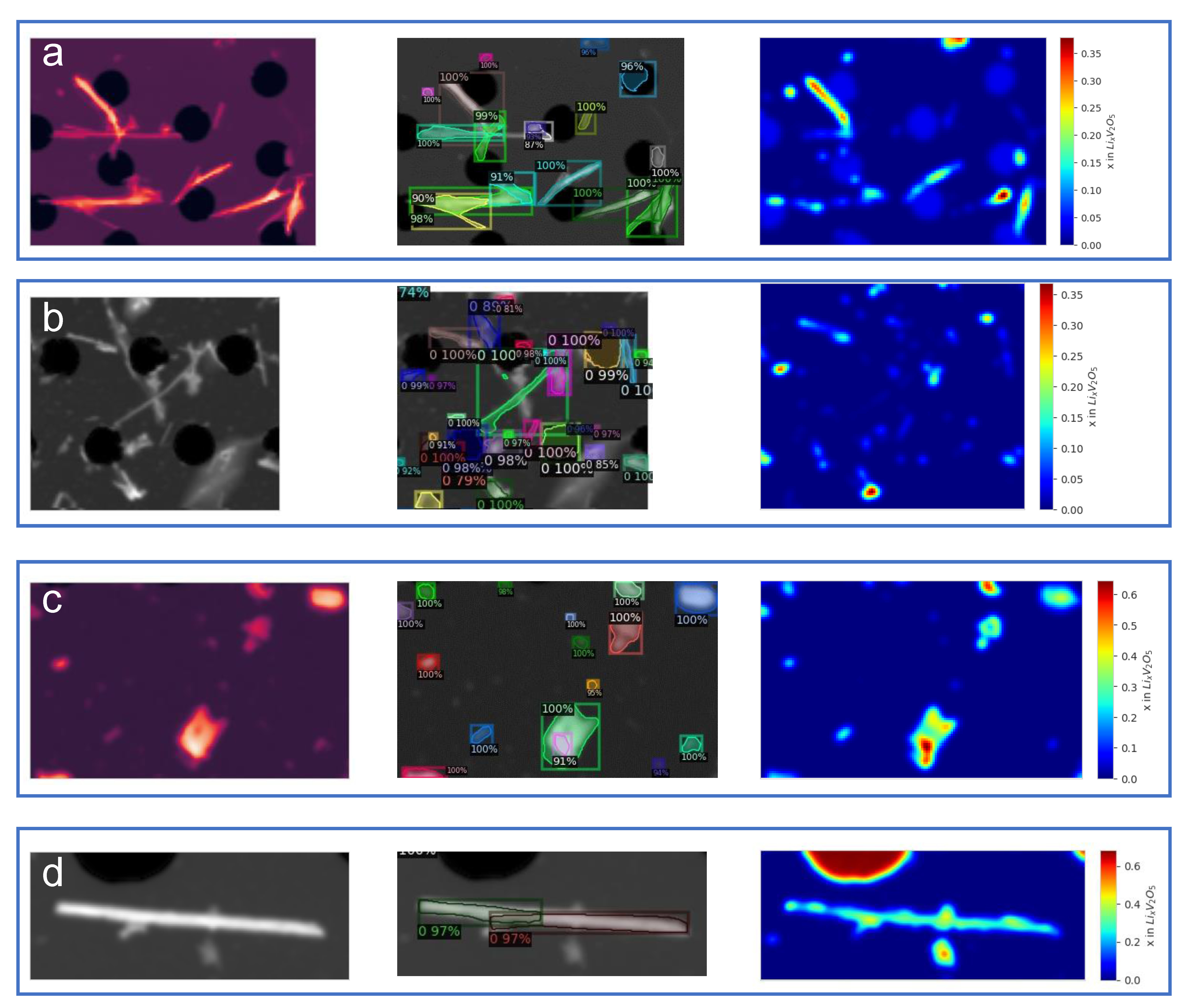}
	\caption{Additional nanoparticle set 2: STXM images (left column), particles after segmentation (middle column), and the lithiation composition maps (right column).}
	\label{fig:set2}
\end{figure}

In Figs.\,\ref{fig:set1} and \ref{fig:set2}, the additional sets of nanoparticles together with the results of segmentation and composition maps other than those investigated in the main text are shown here. These results demonstrate that the proposed deep learning-enabled methodology for particle segmentation and composition maps extraction from STXM images are applicable to particles of dramatically different morphology features. Since the nanoparticle geometry variations in this set of samples are narrow, the geometry-lithiation correlation analysis on it is not feasible and hence is not investigated. 

\bibliographystyle{model3-num-names}